# The number of Euler tours of random directed graphs

Páidí Creed[*][†]  
School of Mathematical Sciences  
Queen Mary, University of London  
p.creed@qmul.ac.uk

Mary Cryan[‡]  
School of Informatics  
University of Edinburgh  
mcryan@inf.ed.ac.uk

**Abstract**

In this paper we obtain the expectation and variance of the number of Euler tours of a random Eulerian directed graph with fixed out-degree sequence. We use this to obtain the asymptotic distribution of the number of Euler tours of a random $d$-in/$d$-out graph and prove a concentration result. We are then able to show that a very simple approach for uniform sampling or approximately counting Euler tours yields algorithms running in expected polynomial time for almost every $d$-in/$d$-out graph. We make use of the BEST theorem of de Bruijn, van Aardenne-Ehrenfest, Smith and Tutte, which shows that the number of Euler tours of an Eulerian directed graph with out-degree sequence $\mathbf{d}$ is the product of the number of arborescences and the term $\frac{1}{n}[\prod_{v \in V}(d_v - 1)!]$. Therefore most of our effort is towards estimating the moments of the number of arborescences of a random graph with fixed out-degree sequence.

## 1 Introduction

### 1.1 Background

Let $G = (V, E)$ be a directed graph. An *Euler tour* of $G$ is any ordering $e_{\pi(1)}, \ldots, e_{\pi(|E|)}$ of the set of arcs $E$ such that for every $1 \leq i < |E|$, the target vertex of arc $e_{\pi(i)}$ is the source vertex of $e_{\pi(i+1)}$, and such that the target vertex of $e_{\pi(|E|)}$ is the source of $e_{\pi(1)}$. We use $ET(G)$ to denote the set of Euler tours of $G$, where two Euler tours are considered to be equivalent if one is a cyclic permutation of the other. It is a well-known fact that a directed graph $G$ has an Euler tour if and only if $G$ is connected and if for each $v \in V$, the in-degree and out-degree of $v$ are equal. In this paper, we are interested in the number of Euler tours of a random Eulerian directed graph with fixed out-degree sequence.

[*]Supported by EPSRC grants EP/F01161X/1 and EP/D043905/1  
[†]Corresponding author  
[‡]Supported by EPSRC grant EP/D043905/1

Let $\mathbf{d} = (d_1, d_2, \ldots)$ be a sequence of positive integers. We let $\mathcal{G}_n^{\mathbf{d}}$ be the space of all Eulerian directed graphs on vertex set $[n] = \{1, 2, \ldots, n\}$ with out-degree sequence $(d_1, d_2, \ldots, d_n)$. We use $m = \sum_{v \in [n]} d_v$ to denote the number of arcs in a graph $G \in \mathcal{G}_n^{\mathbf{d}}$. In the case where $d_i = d_j$ for all $i, j \in [n]$, we refer to the graphs as $d$-in/$d$-out graphs and denote this set by $\mathcal{G}_n^{d,d}$. In this paper, we obtain asymptotic estimates for the first and second moments of the number of Euler tours of a uniformly random $G \in \mathcal{G}_n^{\mathbf{d}}$, for any fixed out-degree vector $\mathbf{d}$.

Using the estimates of the moments, we determine the asymptotic distribution of the number of Euler tours of a random $G \in \mathcal{G}_n^{d,d}$. Similar results have previously been obtained for various structures in the case of *undirected* regular graphs. For example, the asymptotic distribution has already been characterised for Hamiltonian cycles [12, 13, 5], 1-factors [9], and 2-factors [11], in the case of uniformly random $d$-regular undirected graphs. In each of these results, one of the goals was to prove that the structure of interest occurs in $G$ with high probability when $G$ is chosen uniformly at random from the set of all undirected $d$-regular graphs. Since every connected $d$-in/$d$-out graph has an Euler tour, the existence question is not of interest here. However, in the case of Hamiltonian cycles the asymptotic distribution was further used by Frieze et al. [5] to prove that very simple algorithms for random sampling and approximate counting of Hamiltonian cycles run in expected polynomial time for almost every $d$-regular graph. This paper contains analogous counting and sampling results for Euler tours of $d$-in/$d$-out graphs for $d \geq 2$.

Our result uses a well-known relationship between the Euler tours and *arborescences* of an Eulerian graph. An *arborescence* of a directed graph $G = (V, E)$ is a rooted spanning tree of $G$ in which all arcs are directed *towards* the root. We will use $ARBS(G)$ to denote the set of arborescences of $G$ and, for any $v \in V$, use $ARBS(G, v)$ to denote the set of arborescences rooted at $v$. For any Eulerian directed graph $G$, the **BEST** Theorem (due to de **B**ruijn and van Aardenne-**E**hrenfest [17], extending a result of **S**mith and **T**utte [14]) reduces the problem of computing $|ET(G)|$ to the problem of computing $|ARBS(G, v)|$, for any vertex $v \in V$.

**Theorem 1** ([14, 17]). *Let $G = (V, E)$ be an Eulerian directed graph with out-degree sequence $\mathbf{d}$. For any $v \in V$, we have*

$$|ET(G)| = \left[ \prod_{u \in V} (d_u - 1)! \right] |ARBS(G, v)|. \tag{1}$$

The above theorem enables exact counting or sampling of Euler tours of any Eulerian directed graph in polynomial time. For any given digraph $G = (V, E)$, the well-known Matrix-tree theorem shows that for any $v \in V$ the number of arborescences into $v \in V$ exactly equals the value of the $(v, v)$-cofactor of the Laplacian matrix of $G$ (see, for example, [16]). Colbourn et al. [4] gave an algorithm allowing sampling of a random arborescence rooted at $v$ to be carried out in the same time as counting all such arborescences. Hence, applying the **BEST** theorem stated above, the twin tasks of exact counting and uniform sampling of Euler tours of a given Eulerian digraph on $n$ vertices can be performed in the time to evaluate the determinant of an $n \times n$ matrix, which at the time of writing is $O(n^c)$ for $c < 2.3727$[18]. An alternative approach to sampling is presented in [10].

## 1.2 Naïve algorithms

In this paper, we take a different approach and consider a very naïve algorithm for sampling Euler tours of an Eulerian digraph. To describe this algorithm, it helps to introduce the concept of a *transition system* of an Eulerian digraph $G = (V, A)$: for every $v \in V$, consider the set $In(v)$ of arcs directed into $v$, and the set $Out(v)$ of arcs directed away from $v$ (in a multi-graph we allow the possibility that $In(v) \cap Out(v) \neq \emptyset$). We define a pairing $P(v)$ at $v$ to be a matching of $In(v)$ with $Out(v)$. Finally we define a *transition system* of $G$ to be the union of a collection of pairings, one for each vertex of the graph. We let $TS(G)$ denote the set of all transition systems of $G$. If $G$ has the out-degree sequence $d_1, \ldots, d_n$ (for $n = |V|$), then $|TS(G)| = \prod_{i=1}^{n} d_i!$. Note that every Euler tour of $G$ induces a unique transition system on $G$.

Our naïve sampling algorithm presented in Figure 1 generates a random *transition system* for $G$ and tests whether it induces an Euler tour.

**Algorithm** SAMPLE$\langle G = (V, A) \rangle$
**for** $v \in V$ **do**
   Choose a pairing $P(v)$ of $In(v)$ with $Out(v)$, drawn uniformly at random from all pairings.
**end for**
**if** $\cup_{v \in V} P(v)$ induces an Euler tour $T$ on $G$ **then**
   return $T$
**else**
   return $\emptyset$
**end if**

Figure 1: **Algorithm** SAMPLE

We make two simple observations. First, observe that SAMPLE$\langle G = (V, A) \rangle$ generates all transition systems of $G$ with equal probability. Hence all transition systems corresponding to an Euler tour will be generated with a uniform probability (which is $[\prod_{i=1}^{n} d_i!]^{-1}$). Second, the probability that one execution of SAMPLE$\langle G = (V, A) \rangle$ returns an Euler tour is exactly $|ET(G)|/|TS(G)| = |ET(G)| \times [\prod_{i=1}^{n} d_i!]^{-1}$.

In Figure 2, we present our simple approximate counting algorithm. Observe that for any given $\kappa \in \mathbb{N}$, that the expectation $\mathbb{E}[k/\kappa]$ of the value returned by APPROXIMATE$\langle G = (V, A), \kappa \rangle$ is $|ET(G)|/|TS(G)|$. However, the probability that the value returned by APPROXIMATE$\langle G = (V, A), \kappa \rangle$ will be close to $|ET(G)|/|TS(G)|$ depends both on $\kappa$ and on the value of $|ET(G)|$. If we are given a graph $G$ whereby $|ET(G)|$ is guaranteed to be larger than

```
Algorithm APPROXIMATE⟨G = (V, A), κ⟩
k := 0;
for i = 1 → κ do
    T ← SAMPLE⟨G⟩
    if T ≠ ∅ then
        k := k + 1;
    end if
end for
return k/κ
```

Figure 2: **Algorithm** APPROXIMATE

$p(n)^{-1} \prod_{i=1}^{n} d_i!$, where $p(n)$ is some fixed polynomial in $n$, then by setting $\kappa$ appropriately we can guarantee that with high probability APPROXIMATE$\langle G = (V, A), \kappa \rangle$ will return a close approximation of $|ET(G)|/|TS(G)|$. However, there exist Eulerian digraphs where the number of Euler tours is only an exponentially small multiple of $\prod_{i=1}^{n} d_i!$.

In this paper we consider the performance of SAMPLE and of APPROXIMATE on random regular Eulerian digraphs of bounded degree $d$. Our goal will be to show that as the number of vertices grows, that for some $\kappa$ polynomial in $n$, the probability that APPROXIMATE returns a close approximation of $|ET(G)|/|TS(G)|$ tends to 1. This requires that we can demonstrate two things: (i) that the expected number of Euler tours of a random Eulerian digraph of fixed degree is polynomially-related to $|TS(G)| = (d!)^n$; that is, there is some $h > 0$ such that the expected number of Euler tours is greater than $n^{-h}(d!)^n$; (ii) that $|ET(G)|$ on random $d$-regular Eulerian digraphs is concentrated within a window of this expected value.

Note that our algorithms for sampling and approximate counting of random Eulerian digraphs have previously been analysed for the case of Eulerian tournaments in [8]. This was done as part of their analysis of Euler tours on the undirected complete graph with an odd number of vertices. It does not overlap our research - tournaments are regular of degree $(n-1)/2$.

### 1.3 Our proof

The results in this paper are of an asymptotic nature. If $a_n$ and $b_n$ are sequences of numbers, we take $a_n \to b_n$ to mean $\lim_{n \to \infty} a_n/b_n = 1$. Given a sequence of random variables $X_n$ and random variable $Z$, we say $X_n$ *converges in distribution* to $Z$, or $Z$ has the *asymptotic distribution* of $X_n$, if

$$\lim_{n \to \infty} \mathbb{P}[X_n \leq x] = \mathbb{P}[Z \leq x].$$

We generate graphs in $\mathcal{G}_n^{\mathbf{d}}$ using a directed version of the configuration model [2, 3]. We define the configuration space $\Phi_n^{\mathbf{d}}$ as follows. For each $v \in [n]$, let $S_v$ and $T_v$ be disjoint $d_v$-sets and let $S = \cup_{v \in [n]} S_v$ and $T = \cup_{v \in [n]} T_v$. We say $S_v$ is the set of *configuration points* available for arcs leaving $v$ and $T_v$ is the set of points available for arcs entering $v$. A *configuration* $F$ is a perfect matching from $S$ to $T$ and $\Phi_n^{\mathbf{d}}$ is the set of all configurations. Note that $|\Phi_n^{\mathbf{d}}| = m!$. Each configuration $F \in \Phi_n^{\mathbf{d}}$ *projects* to a directed multi-graph $\sigma(F)$ by identifying the elements of $S_v$ and $T_v$. That is, $\sigma(F)$ has an arc $(u, v)$ for each pair from $S_u \times T_v$ that is contained in $F$. This model has been analysed in [1, Section 7], to obtain an estimate of the expected number of Euler tours of a random $G \in \mathcal{G}_n^{\mathrm{d,d}}$ for the case $d = 2$. One nice property of the model, and of the original configuration model, is that directed graphs (without loops or double arcs) are generated with equal probability. Hence, by studying properties of uniformly random configurations it is possible to infer results about uniformly random elements of $\mathcal{G}_n^{\mathbf{d}}$, by conditioning on there being no loops or double arcs.

In Section 2, we consider the configuration model for general (bounded) degree sequences. We first prove a useful combinatorial lemma; then in Theorem 2 we derive and prove exact expressions for the first and second moments for the number of arborescences of $\sigma(F)$, when $F$ is a configuration drawn uniformly at random from $\Phi_n^{\mathbf{d}}$. Next, in Theorem 3, we condition on the event that $\sigma(F)$ is a simple graph, to derive close approximations for the first and second moment, for the number of Arborescences, when $G$ is a simple graph drawn uniformly at random from $\mathcal{G}_n^{\mathbf{d}}$. As an immediate corollary we obtain corresponding approximations for the first and second moment when the random variable is the number of Euler tours. The expected value for the number of Euler tours over $\mathcal{G}_n^{\mathbf{d}}$ is shown in Corollary 1 to tend to the value $\frac{e}{m}(\prod_{v \in [n]} d_v!)$, which is a $\frac{e}{m}$ fraction of $|TS(G)|$. Therefore point (i) of Subsection 1.2 holds.

In the analysis of random structures, it is sometimes the case that we can prove concentration (of a random variable within a fixed range) by applying Chebyshev's inequality to the first and second moment of that random variable. In the final part of Section 2 we show that the values of the first and second moments for Euler Tours in $\mathcal{G}_n^{\mathbf{d}}$ are not good enough to prove concentration of measure using Chebyshev's inequality.

It is for the above reason that in Section 3 we use a more complicated method to show that the number of Euler tours for $G \in \mathcal{G}_n^{\mathbf{d}}$ is asymptotically almost surely close to its expectation. The proof idea we use to obtain an asymptotic distribution is that of *conditioning on short cycle counts*, pioneered by Robinson and Wormald [12, 13]. Implicit in this pair of papers (and the subsequent work of Frieze et al. [5]) is a characterisation of the asymptotic distribution of the number of Hamiltonian cycles in a random $d$-regular graph in terms of random variables counting the number of $i$-cycles, for all fixed positive integers $i$. Janson [6] streamlined the technique of Robinson and Wormald and proved a general theorem (stated by us as Theorem 4). In Section 4, we use Theorem 4 to obtain an asymptotic distribution for the number of Euler tours of a random $d$-in/$d$-out graph.

## 2 Expectation and Variance of Euler tours

In this section, we obtain the expectation and variance of the number of Euler tours of a random $d$-in/$d$-out graph. We will use two particular facts several times in the proofs of this section. Recall the definition of *falling factorial powers*: for every $n, k \in \mathbb{N}$,

$$(n)_k = n(n-1)(n-2)\cdots(n-k+1)\,.$$

**Fact 1.** *Falling factorial powers of sums obey the well known* multinomial theorem

$$(x_1 + x_2 + \cdots + x_l)_k = \sum_{\sum \delta_i = k} \binom{k}{\delta_1, \ldots, \delta_l} \prod_{i=1}^{l} (x_i)_{\delta_i}\,,$$

*where the sum is taken over all partitions of $k$ into $l$ non-negative integer parts.*

**Fact 2** (see, e.g., [15])**.** *Let $V = \{1, 2, \ldots, n\}$, and let $\delta = \{\delta_v : v \in V\}$ be a given vector of non-negative integers. The number of $k$-forests on $V$ in which $v$ has $\delta_v$ children is*

$$\binom{n-1}{k-1}\binom{n-k}{\delta_v : v \in V}\,.$$

We use Fact 1 and Fact 2 to prove the following lemma. In this lemma, and in the proofs of subsequent results, we will speak of a configuration for an (in-directed) arborescence or forest. We take this to mean a partial matching from $S$ to $T$ (in the configuration model) that projects to an arborescence or a forest.

**Lemma 1.** *Suppose we have a set of vertices $V = [n]$ for which there are $x_v$ points for arcs entering $v \in V$ and $y_v$ points for arcs leaving $v \in V$, with $x_v$ not necessarily equal to $y_v$. Then, the number of ways to choose a configuration for an in-directed forest rooted at $R \subseteq V$ is*

$$\left(\prod_{v \in V \setminus R} y_v\right)\left(\sum_{v \in R} x_v\right)\left(\sum_{v \in V} x_v - 1\right)_{n - |R| - 1}\,. \qquad (2)$$

*Proof.* Let $\mathcal{F}$ be a forest on $[n]$ rooted at $R$ and let $\delta_v$ be the number of children of $v$ in $\mathcal{F}$, for each $v \in V$. The number of ways to choose points for the source and target vertex of each arc in $\mathcal{F}$ is

$$\left(\prod_{v \in V \setminus R} y_v\right)\left(\prod_{v \in V} (x_v)_{\delta_v}\right)\,, \qquad (3)$$

since we must choose a point for the start of the arc directed away from each $v \notin R$ and choose one of the $x_v$ points for the end of each of the $\delta_v$ arcs directed towards each $v \in V$.

Let $k = \sum_{v \in R} \delta_v$. We can construct a forest rooted at $R$ by first choosing a $k$-forest on $V \setminus R$, and then attaching each root of this forest as a child of some $v \in R$. The reason we take this approach is to allow us to use Fact 2, which is not explicitly set up to allow us to specifiy particular roots. By Fact 2, the number of $k$-forests on $V \setminus R$ in which $v \in V \setminus R$ has exactly $\delta_v$ children is

$$\binom{n - |R| - 1}{k - 1} \binom{n - |R| - k}{\delta_v : v \in V \setminus R}, \tag{4}$$

and the number of ways to divide the roots of this forest amongst the members of $R$ so that each $v \in R$ has $\delta_v$ children is

$$\binom{k}{\delta_v : v \in R}. \tag{5}$$

Combining (3), (4) and (5) and summing over all possible values for $\delta_v$ gives

$$\left( \prod_{v \in V \setminus R} y_v \right) \times \sum_{k=1}^{n-|R|} \binom{n - |R| - 1}{k - 1} \left( \sum_{(\sum_{v \in R} \delta_v) = k} \binom{k}{\delta_v : v \in R} \prod_{v \in R} (x_v)_{\delta_v} \right)$$
$$\times \left( \sum_{(\sum_{v \in V \setminus R} \delta_v) = n - |R| - k} \binom{n - |R| - k}{\delta_v : v \in V \setminus R} \prod_{v \in V \setminus R} (x_v)_{\delta_v} \right). \tag{6}$$

By Fact 1, we see that the two sums over the different $\delta_v$ in (6) are expansions of the falling factorial powers $(\sum_{v \in R} x_v)_k$ and $(\sum_{v \in V \setminus R} x_v)_{n-|R|-k}$, respectively. Hence, (6) is equal to

$$\left( \prod_{v \in V \setminus R} y_v \right) \sum_{k=1}^{n-|R|} \binom{n - |R| - 1}{k - 1} (\sum_{v \in R} x_v)_k (\sum_{v \in V \setminus R} x_v)_{n-|R|-k}.$$

Applying Fact 1 again gives (2). □

We now use Lemma 1 to analyse the expectation and variance of the number of arborescences in $\sigma(F)$, when $F$ is chosen uniformly at random from $\Phi_n^{\mathbf{d}}$. We say $\mathcal{A} \subset F$ is an arborescence of $F \in \Phi_n^{\mathbf{d}}$ if $\sigma(\mathcal{A})$ is an arborescence of $\sigma(F)$. In the following proofs, we will abuse terminology slightly and switch between speaking of arborescences of configurations and directed graphs arbitrarily. We will define $ARBS(F)$, for any $F \in \mathcal{G}_n^{\mathbf{d}}$, to be the set of partial matchings on $S \times T$ which project to an Arborescence on $[n]$.

**Theorem 2.** *Let $\mathbf{d} = (d_1, d_2, \ldots)$ be a sequence of positive integers. For each $n \in \mathbb{N}$, let $\mathcal{A}_n^\star$ denote the number of arborescences (rooted at any vertex) of a* *uniformly random* $F \in \Phi_n^{\mathbf{d}}$. *Then,*

$$\mathbb{E}[\mathcal{A}_n^\star] = \frac{n}{m}\left[\prod_{v \in [n]} d_v\right];$$

$$\mathbb{E}[(\mathcal{A}_n^\star)^2] = \frac{m}{m-n+1}\mathbb{E}[\mathcal{A}_n^\star]^2.$$

*Proof.* We start by computing the first moment of $\mathcal{A}_n^\star$. To calculate the first moment of $\mathcal{A}_n^\star$ we need to enumerate pairs $(F, \mathcal{A})$, where $F \in \Phi_n^{\mathbf{d}}$ and $\mathcal{A}$ is an arborescence of $F$, and then divide this quantity by $|\Phi_n^{\mathbf{d}}|$. Given $\mathcal{A}$, it is easy to count the number of configurations $F \in \Phi_n^{\mathbf{d}}$ for which $\mathcal{A} \subset F$. In any directed graph $G$ with $m$ arcs, there are exactly $m - n + 1$ arcs not contained in any particular element of $ARBS(G)$. Hence, if we have a configuration for an arborescence, there are $(m - n + 1)!$ ways to extend this to a complete configuration. Applying Lemma 1 with $\mathbf{s} = \mathbf{t} = \mathbf{d}$, we see that the number of arborescences rooted at any particular vertex $v$ is

$$d_v \left(\prod_{u \in [n]\setminus\{v\}} d_u\right)(m-1)_{n-2}. \tag{7}$$

By the BEST theorem (Theorem 1), there are an equal number of arborescences rooted at each vertex of any $F \in \Phi_n^{\mathbf{d}}$. Hence, multiplying (7) by by $n(m-n+1)!$ gives the number of pairs $(F, \mathcal{A})$ with $F \in \Phi_n^{\mathbf{d}}$ and $\mathcal{A} \in ARBS(F)$:

$$n(m-1)!\left(\prod_{v \in [n]} d_v\right).$$

Finally, dividing by the total number of configurations in $\Phi_n^{\mathbf{d}}$, which is $m!$, gives the claimed value for $\mathbb{E}[\mathcal{A}_n^\star]$.

To compute the second moment of $\mathcal{A}_n^\star$ we need to evaluate the following expression

$$\frac{1}{m!}\sum_{F \in \Phi_n^{\mathbf{d}}}|ARBS(F)|^2. \tag{8}$$

We observe that the term $|ARBS(F)|^2$ in (8) is equal to the number of elements in the set

$$\{(\mathcal{A}, \mathcal{A}') : \mathcal{A}, \mathcal{A}' \in ARBS(F)\}.$$

That is,

$$\mathbb{E}[(\mathcal{A}_n^\star)^2] = \frac{|\widetilde{\Phi_n^{\mathbf{d}}}|}{|\Phi_n^{\mathbf{d}}|},$$

where

$$\widetilde{\Phi_n^{\mathbf{d}}} = \{(F, \mathcal{A}, \mathcal{A}') : F \in \Phi_n^{\mathbf{d}}, \mathcal{A}, \mathcal{A}' \in ARBS(F)|\}.$$

Hence, to evaluate $\mathbb{E}[(\mathcal{A}_n^\star)^2]$ we need to count the number of elements of $\widetilde{\Phi_n^{\mathbf{d}}}$.

We compute $|\widetilde{\Phi_n^{\mathbf{d}}}|$ as follows. First, we count the number of ways to choose the intersection of a pair of arborescences $\mathcal{A}$ and $\mathcal{A}'$. Then, we count the number of ways to extend this intersection to $\mathcal{A}$ and $\mathcal{A}'$. Finally, we count the number of ways to choose the remainder of $F$ so that $\mathcal{A}$ and $\mathcal{A}'$ are both in $ARBS(F)$.

We start by considering the final stage. Suppose we have a pair of arborescences $(\mathcal{A}, \mathcal{A}')$ of some configuration $F \in \Phi_n^{\mathbf{d}}$ and suppose $\mathcal{F} = \mathcal{A} \cap \mathcal{A}'$ is a forest rooted at $R \subseteq [n]$. Since we need to add $|R| - 1$ arcs to $\mathcal{F}$ for each arborescence, there will be $n + |R| - 2$ edges in $\mathcal{A} \cup \mathcal{A}'$ and, hence, there are $(m - n - |R| + 2)!$ ways to choose the remaining edges for $F$.

Now we examine the number of different pairs $(\mathcal{A}, \mathcal{A}')$ with $\mathcal{F} = \mathcal{A} \cap \mathcal{A}'$ rooted at $R$. In the analysis that follows, we will overcount slightly, with the pair of Arborescences $(\mathcal{A}, \mathcal{A}')$ depending on the roots of $\mathcal{A}$ and $\mathcal{A}'$. We use the BEST Theorem (Theorem 1) to get back to the correct number at the end of the proof.

We start by counting the number of ways we can choose $\mathcal{F}$, the edges in both arborescences, and then count the number of ways to choose the edges which are in one or the other arborescence. By Lemma 1, the number of ways to choose $\mathcal{F}$ rooted at $R$ is

$$\left(\prod_{v \in [n] \setminus R} d_v\right) \left(\sum_{v \in R} d_v\right) (m-1)_{n-|R|-1}. \tag{9}$$

For each $v \in R$, let $\mathcal{F}_v$ denote the component of $\mathcal{F}$ with root $v$, and let $x_v$ be the number of points in $\bigcup_{u \in \mathcal{F}_v} T_u$ not used by arcs in $\mathcal{F}$. That is,

$$x_v = \sum_{u \in \mathcal{F}_v} d_u - |\mathcal{F}_v| + 1.$$

Note that this is the number of points available to add arcs directed towards vertices of $\mathcal{F}_v$ when we are completing $\mathcal{A}$ and $\mathcal{A}'$. Moreover, we have

$$\sum_{v \in R} x_v = m - n + |R|.$$

We now turn our attention to the number of ways to choose $\mathcal{A} \setminus \mathcal{A}'$ and $\mathcal{A}' \setminus \mathcal{A}$. Choosing the remaining arcs for $\mathcal{A}$ and $\mathcal{A}'$ is equivalent to choosing a pair of disjoint configurations for trees on $R$ in which there are $x_v$ points available for the targets of arcs entering $v$ and $d_v$ points available for the sources of the arcs leaving $v$, for each $v \in R$.

Suppose we have already chosen $\mathcal{A} \setminus \mathcal{A}'$ such that the root of $\mathcal{A}$ is $r$ and suppose that there are $\delta_v$ arcs from $\mathcal{A} \setminus \mathcal{A}'$ directed towards vertices in $\mathcal{F}_v$, for each $v \in R$. Now, suppose we want to choose $\mathcal{A}' \setminus \mathcal{A}$ such that the root of $\mathcal{A}'$ is $r'$, and, for the moment, suppose $r \neq r'$. Choosing $\mathcal{A}' \setminus \mathcal{A}$ amounts to choosing a tree on $R$ rooted at $r'$ in which there are $x_v - \delta_v$ points available for arcs directed towards each $v$, $d_v - 1$ points available for the source of the arc directed away from each $v \neq r$, and $d_r$ points available for the source of the arc directed away from $r$. Hence, by Lemma 1, we see that the number of ways to choose $\mathcal{A}' \setminus \mathcal{A}$ is

$$\frac{(x_{r'} - \delta_{r'})d_r}{(d_r - 1)(d_{r'} - 1)} \left( \prod_{v \in R} (d_v - 1) \right) (m - n)_{|R|-2}.$$

Using Fact 2, we can deduce that the number of ways to choose $\mathcal{A}$ is

$$\frac{\prod_{v \in R} d_v}{d_r} \sum_{\substack{|\boldsymbol{\delta}|=|R|-1 \\ \delta_r \geq 1}} \binom{|R| - 2}{\delta_r - 1; \delta_v : v \in R \setminus \{r\}} \prod_{v \in R} (x_v)_{\delta_v}.$$

Therefore, the number of ways to complete $\mathcal{F}$ to $\mathcal{A} \cup \mathcal{A}'$ is equal to

$$\frac{\prod_{v \in R} d_v (d_v - 1)}{(d_r - 1)(d_{r'} - 1)} (m - n)_{|R|-2}$$

times

$$\sum_{\substack{|\boldsymbol{\delta}|=|R|-1 \\ \delta_r \geq 1}} (x_{r'} - \delta_{r'}) \binom{|R| - 2}{\delta_r - 1; \delta_v : v \in R \setminus \{r\}} \prod_{v \in R} (x_v)_{\delta_v}. \quad (10)$$

We can divide (10) into two sums:

$$x_{r'} \sum_{\substack{|\boldsymbol{\delta}|=|R|-1 \\ \delta_r \geq 1}} \binom{|R| - 2}{\delta_r - 1; \delta_v : v \in R \setminus \{r\}} \prod_{v \in R} (x_v)_{\delta_v} \quad (11)$$

and

$$- \sum_{\substack{|\boldsymbol{\delta}|=|R|-1 \\ \delta_r \geq 1}} \delta_{r'} \binom{|R| - 2}{\delta_r - 1; \delta_v : v \in R \setminus \{r\}} \prod_{v \in R} (x_v)_{\delta_v}. \quad (12)$$

Applying Fact 1, we see that (11) and (12) are equal to $x_r x_{r'}(m-n+|R|-1)_{|R|-2}$ and $-x_r x_{r'}(|R|-2)(m-n+|R|-2)_{|R|-3}$ respectively. Hence, the number of ways to complete $\mathcal{F}$ to $\mathcal{A} \cup \mathcal{A}'$ is

$$\frac{x_r x_{r'}}{(d_r - 1)(d_{r'} - 1)} \left( \prod_{v \in R} d_v(d_v - 1) \right) (m - n + |R| - 2)_{2|R|-4}. \quad (13)$$

If $r = r'$, we can apply an almost identical argument to show that the number of ways to complete $\mathcal{F}$ to $\mathcal{A} \cup \mathcal{A}'$ is

$$\frac{x_r(x_r - 1)}{d_r(d_r - 1)} \left( \prod_{v \in R} d_v(d_v - 1) \right) (m - n + |R| - 2)_{2|R|-4}. \quad (14)$$

Multiplying (13) and (14) by $(d_r - 1)(d_{r'} - 1)$ and $d_r(d_r - 1)$ respectively, and summing over $r$ and $r'$ gives

$$\left( \sum_{r \neq r'} x_r x_{r'} + \sum_r x_r(x_r - 1) \right) \left( \prod_{v \in R} d_v(d_v - 1) \right) (m-n+|R|-2)_{2|R|-4}. \quad (15)$$

Since $\sum_{r\in R} x_r = m - n + |R|$, we have

$$\sum_{r\neq r'} x_r x_{r'} + \sum_r x_r(x_r - 1) = \sum_{r\in R} x_r \left( x_r - 1 + \sum_{r'\neq r}(x_{r'}) \right)$$
$$= (m - n + |R|)(m - n + |R| - 1).$$

Hence, (15) is equal to

$$\left(\prod_{v\in R} d_v(d_v - 1)\right)(m - n + |R|)_{2|R|-2}.$$

Multiplying by the number of ways to choose $\mathcal{F}$, given in (9), and the number of ways to choose the portion of $F$ not contained in $\mathcal{A} \cup \mathcal{A}'$, which is $(m - n - |R| + 2)!$, yields the following expression

$$\left(\prod_{v\in[n]} d_v\right)(m-1)!\left(\prod_{v\in R}(d_v - 1)\right)\left(\sum_{v\in R} d_v\right). \tag{16}$$

The expression (16) over-counts the number of triples $(F, \mathcal{A}, \mathcal{A}')$ in which the intersection $\mathcal{A} \cap \mathcal{A}'$ is a forest rooted at $R$. Each triple $(F, \mathcal{A}, \mathcal{A}')$ in which $\mathcal{A}$ and $\mathcal{A}'$ are rooted at different vertices $u$ and $v$ is counted $(d_u - 1)(d_v - 1)$ times, and each triple $(F, \mathcal{A}, \mathcal{A}')$ in which $\mathcal{A}$ and $\mathcal{A}'$ are rooted at the same vertex $v$ is counted $d_v(d_v - 1)$ times.

Only the second two factors of (16) depend on $R$. Summing these over all $R \subseteq V$ gives

$$\sum_{R\subseteq[n]} \left(\sum_{v\in R} d_v\right)\left(\prod_{v\in R}(d_v - 1)\right), \tag{17}$$

We can evaluate (17) by separating it into $n$ separate sums, each corresponding to the sum over $R \ni v$ for a particular $v \in [n]$,

$$d_v \sum_{R\ni v} \prod_{u\in R}(d_u - 1) = (d_v - 1)\left(\prod_{u\in[n]} d_u\right). \tag{18}$$

Summing the right-hand side of (18) over each $v \in [n]$ and combining with the rest of (16) gives

$$\left(\prod_{v\in[n]} d_v\right)^2 (m - n)(m - 1)!. \tag{19}$$

We cannot immediately obtain the quantity we are looking for from (19) as it over-counts different triples by different amounts. However, by the BEST theorem (Theorem 1), we know that the number of triples $(F, \mathcal{A}, \mathcal{A}')$ in which $\mathcal{A}$ is rooted at $u$ and $\mathcal{A}'$ is rooted at $v$ does not depend on $u$ or $v$, since the projection $\sigma(F)$ is always an Eulerian directed graph. Thus, it follows that the factor by which (19) over-counts the number of triples is

$$\frac{1}{n^2}\left(\sum_{u\neq v}(d_u-1)(d_v-1)+\sum_v d_v(d_v-1)\right) = \frac{(m-n+1)(m-n)}{n^2}. \quad (20)$$

Dividing (19) by (20) and $m!$ gives

$$\mathbb{E}[(\mathcal{A}_n^\star)^2] = \frac{n^2}{m(m-n+1)}\left(\prod_{v\in[n]} d_v\right)^2.$$

$\square$

Recall that simple directed graphs are generated with equal probability in the configuration model. Thus, by conditioning on $\sigma(F)$ containing no loops or 2-cycles, we can obtain the first two moments of the number of arborescences of a uniformly random $G \in \mathcal{G}_n^{\mathbf{d}}$.

**Theorem 3.** *Let $d$ be some fixed constant, let $\mathbf{d} = (d_1, d_2, \ldots)$ be a sequence of positive integers satisfying $d_i \leq d$ for all $i$, let $n \in \mathbb{N}$, and let $m = \sum_{v=1}^n d_v$. Let $\mathcal{A}_n$ denote the number of arborescences of a directed graph chosen randomly from $\mathcal{G}_n^{\mathbf{d}}$. Then, as $m - n \to \infty$,*

$$\mathbb{E}[\mathcal{A}_n] \to e\frac{n}{m}\left[\prod_{v\in[n]} d_v\right];$$
$$\mathbb{E}[\mathcal{A}_n^2] \to e^{-n/m}\frac{m}{m-n}\mathbb{E}[\mathcal{A}_n]^2,$$

*Proof.* In the following we will use $m_2$ to denote $\sum_v d_v^2$.

The proof is as follows. We say $F$ contains a loop at $v$ if there is an edge from $S_v \times T_v$ in $F$ and that $F$ contains a double arc from $u$ to $v$ if there is a pair of edges from $S_u \times T_v$ in $F$. Let $L$ and $D$ denote the number of loops and double arcs in a random $F \in \Phi_n^{\mathbf{d}}$. Then, the event "$F$ is simple" is equivalent to the event $\{L = D = 0\}$. We first analyse the distributions of $L$ and $D$, which we can use to estimate the probability that $F$ is simple. Then, we consider two new random variables, $L^{(1)}$ and $D^{(1)}$, which count the number of loops and double arcs in $F$, when $(F, \mathcal{A})$ is chosen randomly from the set

$$\overline{\Phi_n^{\mathbf{d}}} = \{(F, \mathcal{A}) : F \in \Phi_n^{\mathbf{d}}, \mathcal{A} \in ARBS(F)\} \quad (21)$$

Hence, by analysing the distributions of $L^{(1)}$ and $D^{(1)}$, we can estimate

$$\mathbb{E}[\mathcal{A}_n] = \frac{\mathbb{P}[L^{(1)} = D^{(1)} = 0]}{\mathbb{P}[L = D = 0]}\mathbb{E}[\mathcal{A}_n^\star].$$

Finally, we consider random variables, $L^{(2)}$ and $D^{(2)}$, which count the number of loops and double arcs in $F$, when $(F, \mathcal{A}, \mathcal{A}')$ is chosen randomly from the set

$$\widetilde{\Phi_n^{\mathbf{d}}} = \{(F, \mathcal{A}, \mathcal{A}') : F \in \Phi_n^{\mathbf{d}}, \mathcal{A}, \mathcal{A}' \in ARBS(F)\} \tag{22}$$

Hence, by analysing the distributions of $L^{(2)}$ and $D^{(2)}$, we can estimate

$$\mathbb{E}[(\mathcal{A}_n)^2] = \frac{\mathbb{P}[L^{(2)} = D^{(2)} = 0]}{\mathbb{P}[L = D = 0]} \mathbb{E}[(\mathcal{A}_n^\star)^2].$$

We note that the probability that a random directed graph $G \in \mathcal{G}_n^{\mathbf{d}}$ contains any fixed subgraph $H$ with more arcs than vertices is negligible. To see this, suppose $H$ has $r$ vertices and $r+s$ arcs, where $r$ and $s$ are fixed positive integers. The number of ways to choose a partial configuration projecting to $H$ is $O(n^r)$. However, the probability of a particular set of $r + s$ edges being contained in a uniformly random configuration $F$ is $1/(dn)_{r+s}$. Thus, the probability $\sigma(F)$ contains a subgraph isomorphic to $H$ is $O(n^{-s})$. Then, since the number of different graphs on $r$ vertices with $r+s$ arcs is independent of $n$, we can assume that the contribution to $\mathbb{E}[(L)_j (D)_k]$ from tuples of loops and double arcs with repeated vertices goes to 0 as $n \to \infty$, i.e., for any pair of positive integers $j$ and $k$, we have

$$\mathbb{E}[(L)_j (D)_k] \to \mathbb{E}[L]^j \mathbb{E}[D]^k.$$

Hence, $L$ and $D$ converge to a pair of independent Poisson random variables. The same is true for the random variables $L^{(1)}$ and $D^{(1)}$, and also for $L^{(2)}$ and $D^{(2)}$. Thus, it suffices to compute the means of these random variables to estimate the appropriate probabilities.

We first compute the expectation of $L$ and $D$. Suppose we have a loop edge $e \in S_v \times T_v$ in $F$ and let $I_e$ be the indicator variable for the event $e \in F$. Then, we can write $L = \sum_{v \in V} \sum_{e \in S_v \times T_v} I_e$ and, by linearity of expectation, we have

$$\mathbb{E}[L] = \sum_{v \in V} \sum_{e \in S_v \times T_v} \mathbb{E}[I_e] = \sum_{v \in V} \sum_{e \in S_v \times T_v} \mathbb{P}[e \in F]. \tag{23}$$

Given $e$, the number of ways to choose $F$ with $e \in F$ is $(m-1)!$, so the probability of a random $F \in \Phi_n^{\mathbf{d}}$ containing $e$ is $1/m$. For each $v$, there are $d_v^2$ ways to choose an edge from $S_v \times T_v$. Hence,

$$\mathbb{E}[L] = \frac{1}{m} \sum_v d_v^2 = \frac{m_2}{m}. \tag{24}$$

Next, we compute the expectation of $D$. Here, for every pair of edges $e, f \in S_u \times T_v$, for some $u \neq v$, we define an indicator variable $I_{e,f}$ for the event $e, f \in F$. By linearity of expectation, we have

$$\mathbb{E}[D] = \sum_{u \in V} \sum_{v \in V \setminus \{u\}} \sum_{e, f \in S_u \times T_v} \mathbb{P}[e, f \in F]. \tag{25}$$

The probability of a particular pair of edges $e$ and $f$ occurring in a random configuration $F \in \Phi_n^{\mathbf{d}}$ is, asymptotically, $1/m^2$. Moreover, the number of ways to choose $e, f \in S_u \times T_v$ is $2\binom{d_u}{2}\binom{d_v}{2}$. Hence, the sum in (25) becomes

$$\mathbb{E}[D] \to \frac{2}{m^2} \sum_{u \in V} \sum_{v \in V \setminus \{u\}} \binom{d_u}{2}\binom{d_v}{2}$$

$$= \frac{1}{2m^2} \left( \sum_{u \in V} (d_u)_2 \right)^2 - \frac{1}{2m^2} \sum_{u \in V} (d_u)_2^2 \qquad (26)$$

To finish the calculation we observe that the numerator of the negative term in (26) is $O(m)$ (each $d_u$ is bounded above by a constant $d$, so $\sum_u (d_u)_2^2 \leq d^3 m$). Hence, this part of the sum disappears as $m \to \infty$ and we see that

$$\mathbb{E}[D] \to \frac{(m_2 - m)^2}{2m^2}. \qquad (27)$$

Recall that $L$ and $D$ converge to independent Poisson random variables and, therefore, the probability that $F$ is simple when $F$ is chosen uniformly at random from $\Phi_n^{\mathbf{d}}$ satisfies

$$\mathbb{P}[L = D = 0] \to \exp\left(-\frac{m_2}{m} - \frac{(m_2 - m)^2}{2m^2}\right). \qquad (28)$$

Next, we consider the distributions of $L^{(1)}$ and $D^{(1)}$. We first estimate $\mathbb{E}[L^{(1)}]$. Suppose we have a loop edge $e \in S_v \times T_v$, for some $v \in V$. A loop edge cannot be contained in any arborescence, and, thus, the number of pairs $(F, \mathcal{A}) \in \overline{\Phi_n^{\mathbf{d}}}$ where $e \in F$, is equal to the number of pairs $(F, \mathcal{A}) \in \overline{\Phi_n^{\mathbf{d}'}}$, where $\mathbf{d}'$ is equal to $\mathbf{d}$ with $d_v$ replaced by $d_v - 1$. Hence, from Theorem 2, we can see that the number of elements of $\overline{\Phi_n^{\mathbf{d}}}$ with $e \in F$ is equal to

$$n(d_v - 1) \prod_{u \neq v} d_u (m-2)!. \qquad (29)$$

Dividing (29) by the total number of elements in $\overline{\Phi_n^{\mathbf{d}}}$, which we can also obtain from Theorem 2, gives the probability

$$\mathbb{P}[e \in F | (F, \mathcal{A}) \in \overline{\Phi_n^{\mathbf{d}}}] = \frac{d_v - 1}{d_v(m-1)} \qquad (30)$$

Evaluating (23) with this probability in the place of $\mathbb{P}[e \in F]$ gives

$$\mathbb{E}[L^{(1)}] = \frac{1}{m-1} \sum_v d_v(d_v - 1) \to \frac{m_2 - m}{m}.$$

Next, we evaluate $\mathbb{E}[D^{(1)}]$. Suppose we have a pair of edges $e, f \in S_u \times T_v$ for some $u \neq v$. By Lemma 1, the number of arborescences rooted at $u$ in which each $w \notin \{u, v\}$ has $d_w$ points available for its incoming and outgoing arcs, $u$ has $d_u$ points available for incoming arcs, and $v$ has $d_v - 2$ points available for incoming arcs and $d_v$ available for outgoing arcs is

$$\left(\prod_{w=1}^{n} d_w\right) (m-3)_{n-2}. \tag{31}$$

The expression in (31) counts the number of partial configurations which consist of the edges $e$ and $f$ along with $n-1$ configuration edges that project to an arborescence rooted at $u$. There are $(m-n-1)!$ ways to extend each of these partial configurations to some $F \in \Phi_n^{\mathbf{d}}$. Hence, the following expression counts the number of pairs $(F, \mathcal{A}) \in \overline{\Phi_n^{\mathbf{d}}}$ with $e, f \in F$ and $\mathcal{A}$ rooted at $u$.

$$\left(\prod_{w=1}^{n} d_w\right) (m-3)!. \tag{32}$$

By the BEST Theorem (Theorem 1), we know that each $F \in \Phi_n^{\mathbf{d}}$ has the same number of arborescences rooted at each vertex, so (32) counts exactly $1/n$ of the pairs $(F, \mathcal{A}) \in \overline{\Phi_n^{\mathbf{d}}}$ with $e, f \in F$. Multiplying (32) by $n$ and dividing by $|\overline{\Phi_n^{\mathbf{d}}}|$ gives

$$\mathbb{P}[e, f \in F | (F, \mathcal{A}) \in \overline{\Phi_n^{\mathbf{d}}}] \to \frac{1}{m^2}. \tag{33}$$

This is the same probability as when $F$ is chosen uniformly at random from $\Phi_n^{\mathbf{d}}$, so evaluating (27) with (33) in place of $\mathbb{P}[e, f \in F]$ does not change the (asymptotic) value and we have

$$\mathbb{E}[D^{(1)}] \to \mathbb{E}[D].$$

Again, since $L^{(1)}$ and $D^{(1)}$ converge to independent Poisson random variables we have that the probability of $F$ being simple in a random $(F, \mathcal{A}) \in \overline{\Phi_n^{\mathbf{d}}}$ satisfies

$$\mathbb{P}[L^{(1)} = D^{(1)} = 0] \to \exp\left(-\frac{m_2 - m}{m} - \frac{(m_2 - m)^2}{2m^2}\right). \tag{34}$$

Together (28) and (34) give the claimed estimate for $\mathbb{E}[\mathcal{A}_n]$.

Finally, we consider the distributions of $L^{(2)}$ and $D^{(2)}$. First, suppose we have a loop edge $e \in S_v \times T_v$. The number of elements of $\widetilde{\Phi_n^{\mathbf{d}}}$ with $e \in F$ is equal to the number of elements of $\widetilde{\Phi_n^{\mathbf{d}'}}$, where $\mathbf{d}'$ is the out-degree vector we used to compute $\mathbb{E}[L^{(1)}]$. By Theorem 2, we have

$$|\widetilde{\Phi_n^{\mathbf{d}'}}| = \frac{(d_v - 1)^2}{(d_v)^2} \frac{n^2}{m-n} \left(\prod_{w \in V} d_w\right)^2 (m-2)!.$$

Dividing by the number of elements in $\widetilde{\Phi_n^{\mathbf{d}}}$, which know from Theorem 2, we see that

$$\mathbb{P}[e \in F | (F, \mathcal{A}, \mathcal{A}') \in \widetilde{\Phi_n^{\mathbf{d}}}] \to \frac{(d_v - 1)^2}{(d_v)^2 m}.$$

Evaluating (24) with this probability in the place of $\mathbb{P}[e \in F]$ gives

$$\mathbb{E}[L^{(2)}] \to \frac{m_2 - 2m + n}{m}. \tag{35}$$

We now evaluate $\mathbb{E}[D^{(2)}]$. Suppose we have a pair of edges $e, f \in S_u \times T_v$ for some $u \neq v$. There are three cases to consider: $e, f \in \mathcal{A} \cup \mathcal{A}'$; $e, f \notin \mathcal{A} \cup \mathcal{A}'$; or exactly one of $e$ and $f$ is in $\mathcal{A} \cup \mathcal{A}'$. We estimate $\mathbb{E}[D^{(2)}]$ as follows. Using slightly more general arguments than those used to compute the second moment in Theorem 2, we count the number of triples $(F, \mathcal{A}, \mathcal{A}')$ for each of these three cases, obtaining expressions which overcount in the same way as (19). Then, since the way in which triples are over-counted is the same in each of the three analyses, i.e., the number of times each triple $(F, \mathcal{A}, \mathcal{A}')$ is counted is determined by the out-degrees of the roots of $\mathcal{A}$ and $\mathcal{A}'$, we can add these three expressions together, apply the BEST theorem, and proceed as we did in the proof of Theorem 2.

In each of the three cases, we want to count pairs of arborescences using some subset of the configuration points. Suppose we are working with sets of points where $s_v = |S_v|$ and $t_v = |T_v|$ for each $v$, with $s_v$ not necessarily equal to $t_v$. Note that the fact that the in-degree and out-degree of each vertex $v$ are equal is only used at the last step of the analysis of the second moment of $\mathcal{A}_n^\star$ (in Theorem 2). Thus, by following the arguments of the second part of Theorem 2 we find that, for each $R \subseteq V$, the expression over-counting triples $(F, \mathcal{A}, \mathcal{A}')$ where $\mathcal{A} \cap \mathcal{A}'$ is a forest rooted at $R$ (given by (16) in the proof of Theorem 2) becomes

$$\left(\sum_{w \in R} t_w\right) \left(\prod_{w \in R} (s_w - 1)\right) \left(\prod_{w \in V} s_w\right) (m-1)!. \tag{36}$$

The factor by which (36) over-counts $(F, \mathcal{A}, \mathcal{A}')$ is $(s_r - 1)(s_{r'} - 1)$ if $\mathcal{A}$ and $\mathcal{A}'$ are rooted at different vertices $r, r' \in R$, and is $s_r(s_r - 1)$ if both are rooted at the same vertex $r \in R$. Summing (36) over all possibilities for $R$ gives

$$\left(\sum_{w \in V} \frac{t_w(s_w - 1)}{s_w}\right) \left(\prod_{w \in V} s_w\right)^2 (m-1)! \tag{37}$$

Now, suppose $e, f \notin \mathcal{A} \cup \mathcal{A}'$. To enumerate the number of triples of this form we evaluate (37) with $s_w = d_w$ for $w \neq u$, $s_u = d_u - 2$, $t_w = d_w$ for $w \neq v$, and $t_v = d_v - 2$, since we are removing two points from each of $S_u$ and $T_v$. This gives

$$\left(m - n - \frac{d_u}{d_u - 2} - \frac{d_v - 2}{d_v}\right) \frac{(d_u - 2)^2}{d_u^2} \left(\prod_{w \in V} d_w\right)^2 (m-3)!,$$

or, asymptotically, as $m - n \to \infty$,

$$(m-n) \frac{(d_u - 2)^2}{d_u^2} \left(\prod_{w \in V} d_w\right)^2 (m-3)!, \tag{38}$$

Next, suppose $e, f \in \mathcal{A} \cup \mathcal{A}'$. Since there can be at most one arc leaving $u$ in $\mathcal{A}$ or $\mathcal{A}'$ it follows that we have an arc $(u, v)$ in both $\mathcal{A}$ and $\mathcal{A}'$. Hence, when we are choosing the pair of arborescences we must assume that $(u, v)$ is always present. The corresponds to replacing $u$ and $v$ by a single vertex $v'$ which has $d_v$ points available for outgoing arcs and $d_u + d_v - 2$ points available for incoming arcs. That is, in this instance we work in a model where $s_w = t_w = d_w$ for $w \notin \{u, v\}$, $s_{v'} = d_v$, and $t_{v'} = d_u + d_v - 2$. Evaluating (37) with these values yields, asymptotically,

$$(m-n)\frac{1}{(d_u)^2}\left(\prod_{w \in V} d_w\right)^2 (m-3)!. \tag{39}$$

Given a pair of points from each of $S_u$ and $T_v$, there are two ways to choose $e$ and $f$ and two ways to assign then to $\mathcal{A}$ and $\mathcal{A}'$. Thus, the number of triples satisfying $e, f \in \mathcal{A} \cup \mathcal{A}'$ is, asymptotically,

$$(m-n)\frac{4}{(d_u)^2}\left(\prod_{w \in V} d_w\right)^2 (m-3)!. \tag{40}$$

Finally, suppose exactly one of $e$ and $f$ is in $\mathcal{A} \cup \mathcal{A}'$. This case is a little bit more complicated. Suppose $e \in \mathcal{A}$. We contract $u$ and $v$ to a single vertex $v'$, as in the previous case, and choose a forest on $V \setminus \{u, v\}$ and $v'$ with root $R$. Then, we proceed as in the proof of Theorem 2, except we consider $u$ to be one of the roots of the components of $\mathcal{F}$ when choosing $\mathcal{A}'$; that is, when choosing $\mathcal{A} \setminus \mathcal{A}'$ we choose a tree on $R$, but when choosing $\mathcal{A}' \setminus \mathcal{A}$ we choose a tree on $R \cup \{u\}$, where the vertices in the component of our initial forest have now been divided between a component rooted at $u$ and a component containing $v$. The final expression counting the number of ways to complete $\mathcal{F}$ to $\mathcal{A} \cup \mathcal{A}'$ does not depend on how the vertices are distributed amongst the components of $\mathcal{F}$, so it is safe to do this. In this way, we obtain the expression

$$(d_u - 2)\left(\sum_{w \in R} t_w\right)\left(\prod_{w \in R}(s_w - 1)\right)\left(\prod_{w \in V \setminus \{u\}} s_w\right)(m-1)! \tag{41}$$

which over-counts the number of triples $(F, \mathcal{A}, \mathcal{A}')$ where $R \cup \{u\}$ are the roots of the components of $\mathcal{A} \cap \mathcal{A}'$, $e \in \mathcal{A}$, and $f \in F \setminus (\mathcal{A} \cup \mathcal{A}')$.

Proceeding as before, by summing over all possibilities for $R$, gives

$$(m-n)\frac{d_u - 2}{(d_u)^2}\left(\prod_{w \in V} d_w\right)^2 (m-1)!.$$

The cases where $e \in \mathcal{A}'$, $f \in \mathcal{A}$, and $f \in \mathcal{A}'$ are all equivalent, so the expression over-counting triples $(F, \mathcal{A}, \mathcal{A}')$ with exactly one of $e$ and $f$ in $\mathcal{A} \cup \mathcal{A}'$ is

$$(m-n)\frac{4(d_u - 2)}{(d_u)^2}\left(\prod_{w \in V} d_w\right)^2 (m-1)!. \tag{42}$$

Adding (38), (40), and (42) gives

$$(m - n)\left(\prod_{w \in V} d_w\right)^2 (m-3)!. \tag{43}$$

Finally, we observe that the triples are over-counted consistently in the three separate constructions given above. Hence, we can conclude, by the same reasoning as was used in Theorem 2, that (43) over-counts the elements of $\widetilde{\Phi_n^{\mathbf{d}}}$ by a factor of

$$\frac{(m-n-1)(m-n-2)}{n^2}. \tag{44}$$

Dividing (43) by (44) and the number of elements in $\widetilde{\Phi_n^{\mathbf{d}}}$ gives

$$\mathbb{P}[e, f \in F | (F, \mathcal{A}, \mathcal{A}') \in \widetilde{\Phi_n^{\mathbf{d}}}] \to \frac{1}{m^2}.$$

This is the same probability for $e, f \in F$ when $F$ is chosen uniformly at random from $\Phi_n^{\mathbf{d}}$, so we can conclude

$$\mathbb{E}[D^{(2)}] \to \mathbb{E}[D].$$

The random variables $L^{(2)}$ and $D^{(2)}$ converge to independent Poisson random variables. Hence, the probability that $F$ is simple, when $(F, \mathcal{A}, \mathcal{A}')$ is chosen uniformly at random from $\widetilde{\Phi_n^{\mathbf{d}}}$, satisfies

$$\mathbb{P}[L^{(2)} = D^{(2)} = 0] \to \exp\left(-\frac{m_2 - 2m + n}{m} - \frac{(m_2 - m)^2}{2m^2}\right) \tag{45}$$

Combining (28) and (45) gives the claimed estimate for $\mathbb{E}(\mathcal{A}_n)^2$. $\square$

Given the expectation and variance of the number of arborescences of a random $G \in \mathcal{G}_n^{\mathbf{d}}$, we can, from the BEST Theorem (Theorem 1), deduce the expectation and variance of the number of Euler tours of a uniformly random $G \in \mathcal{G}_n^{\mathbf{d}}$.

**Corollary 1.** *Let $d$ be some fixed constant, let $\mathbf{d} = (d_1, d_2, \ldots)$ be a sequence of positive integers satisfying $d_i \leq d$ for all $i$, let $n \in \mathbb{N}$, and let $m = \sum_{v=1}^n d_v$. Let $\mathcal{T}_n$ denote the number of Euler tours of a directed graph chosen randomly from $\mathcal{G}_n^{\mathbf{d}}$. Then, as $m - n \to \infty$,*

$$\mathbb{E}[\mathcal{T}_n] \to \frac{e}{m}\left[\prod_{v \in [n]} (d_v)!\right];$$

$$\mathbb{E}[\mathcal{T}_n^2] \to e^{-n/m}\frac{m}{m-n}\mathbb{E}[\mathcal{T}_n]^2.$$

We now consider our estimates for the first and second moment in the context of Chebyshev's inequality, which states that for a random variable $X$ with expectation $\mu(X)$ and variance $\sigma(X)$, that for any $k > 0$, the probability that $X$ deviates by more than $k\sigma(X)$ from its mean is bounded as follows:

$$\mathbb{P}[|X - \mathbb{E}(X)| \geq k\sigma(X)] \leq \frac{1}{k^2}$$

For us $X$ is $\mathcal{T}_n$, hence

$$\sigma(\mathcal{T}_n) = [\mathbb{E}[\mathcal{T}_n^2] - \mathbb{E}[\mathcal{T}_n]^2]^{1/2}$$

$$\sim \mathbb{E}[\mathcal{T}_n]\left[e^{-n/m}\frac{m}{m-n} - 1\right]^{1/2}$$

$$\sim \mathbb{E}[\mathcal{T}_n]\left[\frac{m}{m-n}((1-\frac{n}{m}) + \frac{1}{2}(\frac{n}{m})^2(1-\frac{n}{3m})) - 1\right]^{1/2}$$

$$= \mathbb{E}[\mathcal{T}_n]\left[1 + \frac{1}{2}\frac{n^2}{m(m-n)}(1-\frac{n}{3m}) - 1\right]^{1/2}$$

$$= \mathbb{E}[\mathcal{T}_n]\left[\frac{n^2}{2m(m-n)}(1-\frac{n}{3m})\right]^{1/2}$$

$$= \mathbb{E}[\mathcal{T}_n]\frac{n}{\sqrt{2m}}\left[\frac{3m-n}{3(m-n)}\right]^{1/2}$$

Assume $\widehat{d} = (\sum_{v \in [n]} d_v)/n$, and note we have $2 \leq \widehat{d} \leq d$. Then we have $\sigma(\mathcal{T}_n) \sim \mathbb{E}[\mathcal{T}_n]\frac{1}{\sqrt{2\widehat{d}}}(\frac{\widehat{d}-1/3}{\widehat{d}-1})^{1/2} \geq 2^{-1/2}\widehat{d}^{-1}$. Therefore, if we consider any $k > \sqrt{2\widehat{d}}$ in Chebyshev's inequality, we will have $k\sigma(\mathcal{T}_n) > \mathbb{E}[\mathcal{T}_n]$. In this case, Chebyshev's inequality covers a range of values for $\mathcal{T}_n$ as small as 0, which is not helpful for our analysis. Hence the only values of $k$ which can potentially help show $\mathcal{T}_n$ is not too small are $k \leq \sqrt{2\widehat{d}}$. However, this forces $k$ to be upper bounded by the constant $\sqrt{2d}$, meaning that the probability of the favourable outcome can be no greater than $1 - \frac{1}{\sqrt{2}}d$, which does not approach 1. Hence Chebyshev's Inequality cannot give the desired results.

In the next section, we will show how to use results of this section and the estimates of Corollary 1 to obtain an asymptotic distribution for the number of Euler tours of a random $G \in \mathcal{G}_n^{d,d}$, from which we can derive a concentration inequality.

## 3 Asymptotic distribution of Euler tours

To compute the asymptotic distribution we will use the following general theorem of Janson [6] (see also [7, Chapter 9]).

**Theorem 4** (Janson [6]). *Let $\lambda_i > 0$ and $\delta_i \geq -1$, $i = 1, 2, \ldots$, be constants and suppose that for each $n$ there are random variables $X_{in}$, $i = 1, 2, \ldots$, and $Y_n$ (defined on the same probability space) such that $X_{in}$ is non-negative integer valued and $\mathbb{E}[Y_n] \neq 0$ (at least for large $n$) and furthermore the following conditions are satisfied:*

1. *$X_{in} \to X_{i\infty}$ (in distribution) as $n \to \infty$, jointly for all $i$, where $X_{i\infty}$ is a Poisson random variable with mean $\lambda_i$;*

2. *For any finite sequence $x_1, \ldots x_k$ of non-negative integers*

$$\frac{\mathbb{E}[Y_n | X_{1n} = x_1, \ldots X_{kn} = x_k]}{\mathbb{E}[Y_n]} \to \prod_{i=1}^{k} (1 + \delta_i)^{x_i} e^{-\lambda_i \delta_i} \quad as \ n \to \infty;$$

3. *$\sum_i \lambda_i \delta_i^2 < \infty$;*

4. *$\frac{\mathbb{E}[Y_n^2]}{\mathbb{E}[Y_n]^2} \to \exp(\sum_i \lambda_i \delta_i^2)$;*

*Then*

$$\frac{Y_n}{\mathbb{E}[Y_n]} \to W = \prod_{i=1}^{\infty} (1 + \delta_i)^{X_{i\infty}} e^{-\lambda_i \delta_i}.$$

*Moreover, this and the convergence in (1) holds jointly. The infinite product defining $W$ converges a.s. and in $L_2$, with $\mathbb{E}[W] = 1$ and $\mathbb{E}[W^2] = \exp(\sum_i \lambda_i \delta_i^2) = \lim_{n \to \infty} \mathbb{E}[Y_n]^2 / \mathbb{E}[Y_n]^2$. Hence, the normalised variables are uniformly square integrable. Furthermore, the event $W = 0$ equals, up to a set of probability $0$, the event that $X_{i\infty} > 0$ for some $i$ with $\delta_i = -1$. In particular, $W > 0$ a.s. if and only if every $\delta_i > -1$.*

In our application of Theorem 4 we will have $Y_n = \mathcal{T}_n$, the random variable counting Euler tours of $d$-in/$d$-out graphs, and $X_{in}$ equal to the number of directed $i$-cycles in a random $d$-in/$d$-out graph. To apply Theorem 4 we need the following two lemmas.

**Lemma 2.** *For each positive integer $i$ let $X_{in}$ count the number of directed $i$-cycles in a directed graph obtained as the projection of a uniformly random $F \in \Phi_n^{\mathbf{d}}$. The variables $X_{in}$ are asymptotically independent Poisson random variables with means $\mathbb{E}[X_{in}] = \lambda_i = \frac{d^i}{i}$.*

*Proof.* Recall that the probability of a uniformly random $G \in \mathcal{G}_n^{\mathbf{d}}$ containing any particular fixed subgraph $H$ with more arcs than vertices is negligible. Hence, we can assume that cycles occur independently, i.e., for any sequence of non-negative integers $k_1, \ldots, k_\ell$, we have

$$\mathbb{E}\left[\prod_{i=1}^{\ell} (X_{in})_{k_i}\right] \to \prod_{i=1}^{\ell} \lambda_i^{k_i}.$$

Hence, the random variables $X_{1n}, \ldots, X_{\ell n}$ converge to a set of independent Poisson random variables. Thus, all that remains is to estimate $\mathbb{E}[X_{in}]$.

We say a set of $i$ edges $e_1, e_2, \ldots, e_i$ in a configuration is an $i$-cycle if there is a sequence of distinct vertices $v_1, v_2, \ldots, v_i$ such that $e_j \in S_{v_j} \times T_{v_{j+1}}$ for $j < i$ and $e_i \in S_{v_i} \times T_{v_1}$. The probability of any particular $i$-cycle being contained in a random $F \in \Phi_n^{\mathbf{d}}$ is

$$\frac{(dn-i)!}{(dn)!} \to \frac{1}{(dn)^i}.$$

So, to estimate $\mathbb{E}[X_{in}]$ all we need to do is count the number of different $i$-cycles that can occur in some $F \in \Phi_n^{\mathbf{d}}$ and then divide by $(dn)^i$. Let $I$ be some $i$-subset of $[n]$. There are $(i-1)!$ different ways to arrange $I$ into an $i$-cycle $(v_1, v_2, \ldots, v_i)$ and then $d^{2i}$ ways to choose edges $e_j \in S_{v_j} \times T_{v_{j+1}}$ for $1 \leq j < i$ and $e_i \in S_{v_i} \times T_{v_1}$. Hence,

$$\mathbb{E}[X_{in}] \to \frac{(i-1)!}{(dn)^i} \binom{n}{i} d^{2i},$$

and so $\mathbb{E}[X_{in}] \to \lambda_i$. $\square$

**Lemma 3.** *Let $X_{in}$ be as in Lemma 3 and let $\mu_i = \frac{d^i - 1}{i}$. Then, for any fixed set of integers $k_1, k_2, \ldots, k_\ell$ we have*

$$\frac{\mathbb{E}[\mathcal{A}_n^\star \prod_{i=1}^\ell (X_{in})_{k_i}]}{\mathbb{E}[\mathcal{A}_n^\star]} \to \prod_{i=1}^\ell \mu_i^{k_i}.$$

*Proof.* We only verify

$$\frac{\mathbb{E}[\mathcal{A}_n^\star X_{in}]}{\mathbb{E}[\mathcal{A}_n^\star]} \to \mu_i;$$

convergence of the factorial moments holds for the same reasons as were given in Lemma 2.

Let $\overline{\Phi}_n^{d,d}$ be the set defined in (21) (for the case $d_v = d$ for all $v$) and let $I$ be an $i$-subset of $[n]$. As in the previous lemma, there are $(i-1)!d^{2i}$ ways to choose a configuration for an $i$-cycle on $I$. To estimate $\mathbb{E}[\mathcal{A}_n^\star X_{in}]$ we need to calculate the probability that a particular $i$-cycle $C$ is contained in $F$ when $(F, \mathcal{A})$ is chosen uniformly at random from $\overline{\Phi}_n^{d,d}$. Suppose $C \cap \mathcal{A}$ has $c$ components, $P_1, P_2, \ldots, P_c$, each of which is a directed path, and let $v_j$ be the final vertex in the path $P_j$ for $1 \leq j \leq c$. Choosing the remainder of $\mathcal{A}$ is then equivalent to choosing an arborescence on $(V \setminus I) \cup \{v_j : 1 \leq j \leq c\}$, where we have collapsed each path to a single vertex. Each $v \in V \setminus I$ has $d$ points available for arcs directed towards or away from $v$. For each $j = 1 \ldots c$, there are $|P_j|(d_u - 1)$ points available for arcs directed towards $v_j$, and $d-1$ points available for arcs directed away from $v_j$. Once we have chosen $\mathcal{A}$, there are $(dn - n - c - 1)!$ ways to complete $F$. Hence, using Lemma 1, we can deduce that the number of ways to choose the remainder of $(F, \mathcal{A})$ is

$$n(d-1)^c d^{n-i}(dn-i-1)!.$$

Summing over all the possible choices for $P = \{v_1, v_2, \ldots, v_c\}$, dividing by $|\overline{\Phi_n^{d,d}}| = d^{n-1}(dn)!$, and applying Stirling's formula yields

$$\mathbb{E}[\mathcal{A}_n^\star X_{in}] \to \mu_i \mathbb{E}[\mathcal{A}_n^\star].$$

$\square$

**Corollary 2.** *Let $d \geq 2$ be some fixed integer, and let $\mathcal{T}_n$ denote the number of Euler tours in a directed graph $G$ chosen uniformly at random from $\mathcal{G}_n^{d,d}$. For any fixed set of integers $k_2, \ldots, k_\ell$ we have*

$$\frac{\mathbb{E}[\mathcal{T}_n \prod_{i=1}^\ell (X_{in})_{k_i}]}{\mathbb{E}[\mathcal{T}_n]} \to \prod_{i=2}^\ell \mu_i^{k_i}.$$

We are now able to apply Janson's theorem to obtain an asymptotic distribution for the number of Euler tours of a uniformly random $G \in \mathcal{G}_n^{d,d}$.

**Theorem 5.** *Let $d \geq 2$ be some fixed integer, and let $\mathcal{T}_n$ denote the number of Euler tours in a directed graph $G$ chosen uniformly at random from $\mathcal{G}_n^{d,d}$. Then,*

$$\frac{\mathcal{T}_n}{\mathbb{E}[\mathcal{T}_n]} \to \prod_{i=2}^\infty \left(1 - \frac{1}{d^i}\right)^{Z_i} e^{1/i},$$

*where the $Z_i$ are independent Poisson random variables with means $d^i/i$.*

*Proof.* It suffices to show that conditions (1) to (4) of Theorem 4 are satisfied by $\mathcal{T}_n$ and $\{X_{in} : i \geq 2\}$, where $X_{in}$ is the random variable counting $i$-cycles. Lemma 2 and Corollary 2 provide conditions (1) and (2) with

$$\lambda_i = \frac{d^i}{i} \quad \text{and} \quad \delta_i = -\frac{1}{d^i}.$$

With these values, evaluating the sum in condition (3) gives

$$\sum_{i=2}^\infty \frac{1}{id^i} = -\frac{1}{d} + \log\left(\frac{d}{d-1}\right). \tag{46}$$

Finally, Corollary 1 provides condition (4). $\square$

## 4 Generating and counting Euler tours

We now turn to the analysis of Algorithm SAMPLE in Section 1. Note that although the algorithm is defined in terms of transition systems, it can also be considered as equivalent to a random directed walk on the Eulerian digraph or graph; terminating when we have used all outgoing edges of the starting vertex (whether we have created an Euler tour or not). This procedure was first considered in [8], where the authors considered it for undirected graphs and showed that the expected number of runs needed to obtain an Euler tour is polynomial for the case of $G = K_n$ for odd $n$. We consider the algorithm for $G \in \mathcal{G}_n^{d,d}$ and are interested in the quantitity

$$\frac{|ET(G)|}{(d!)^n}. \tag{47}$$

The following theorem uses the results of the previous section (by a similar argument to that used in [5, Lemma 1]) to show that this value is $\Omega(n^{-2})$ with high probability when $G$ is chosen uniformly at random from $\mathcal{G}_n^{d,d}$. When this is the case, we can generate uniformly random Euler tours of $G$ in expected polynomial time. Moreover, by setting the value of $\kappa$ in APPROXIMATE appropriately, we can approximate $|ET(G)|$.

**Theorem 6.** *Let $d$ be some fixed integer, $d \geq 2$, and let $G$ be chosen uniformly at random from $\mathcal{G}_n^{d,d}$. Then,*

$$\mathbb{P}\left[\frac{|ET(G)|}{(d!)^n} \in \Omega(n^{-2})\right] \to 1,$$

*as $n \to \infty$.*

*Proof.* We first note that by the estimate for $\mathbb{E}[\mathcal{T}_n]$ given in Corollary 1, the statement above is equivalent to showing that

$$\mathbb{P}\left[\mathcal{T}_n \geq n^{-1}\mathbb{E}[\mathcal{T}_n]\right] \to 1.$$

For $\mathbf{x} = (x_2, \ldots, x_k)$ we define $\mathcal{G}_\mathbf{x}$ to be the set of all $d$-in/$d$-out graphs containing exactly $x_i$ directed cycles of length $i$ for each $i = 2 \ldots k$, and

$$W(\mathbf{x}) = \prod_{i=2}^{k}\left(1 - \frac{1}{d^i}\right)^{x_i} e^{1/i}.$$

For each fixed $\gamma > 0$ we define

$$S(\gamma) = \{\mathbf{x} : x_i \leq \lambda_i + \gamma\lambda_i \text{ for } 2 \leq i \leq k\}.$$

From Lemma 2 (and Lemma 3 of [5]), we can deduce that the probability of a random $d$-in/$d$-out graph $G$ not being contained in $\mathcal{G}_\mathbf{x}$ for some $\mathbf{x} \in S(\gamma)$ is $O(e^{-a\gamma})$, where $a$ is an absolute constant independent of $\gamma$. Hence, to verify the theorem all we need do is show that

$$W(\mathbf{x}) \geq e^{-(b+c)\gamma} \quad \forall \mathbf{x} \in S(\gamma), \tag{48}$$

where $b$ and $c$ are absolute constants independent of $\gamma$. For any particular $b, c$ and $\gamma$, we can choose $n$ sufficiently large so that $e^{-(b+c\gamma)} \geq n^{-1}$. Then, if (48) holds, we have

$$\mathbb{P}\left[\mathcal{T}_n \geq n^{-1}\mathbb{E}[\mathcal{T}_n]\right] \geq 1 - e^{-a\gamma}.$$

The above holds for any constant $\gamma$, and so can be taken as equal to 1 in the limiting case.

So, it remains to prove (48). For $x \in S(\gamma)$ we have $W(\mathbf{x}) = AB^\gamma$, where

$$A = \prod_{i \geq 2} \left(1 - \frac{1}{d^i}\right)^{\lambda_i} e^{1/i} \qquad (49)$$

$$B = \prod_{i \geq 2} \left(1 - \frac{1}{d^i}\right)^{\lambda_i^{2/3}}. \qquad (50)$$

We can bound the right hand side of (49) as

$$A \geq \prod_{i=2}^\infty \exp\left(\frac{1}{i} - \frac{d^i}{i(d^i - 1)}\right) = \exp\left(\sum_{i=2}^\infty -\frac{1}{i(d^i - 1)}\right).$$

The sum inside the exponential is clearly convergent, so we can conclude that $A \geq e^{-b}$ for some absolute constant $b$. Similarly, we can bound $B$ by

$$B \geq \exp\left(-\sum_{i=2}^\infty \frac{1}{(i)^{2/3} d^{i/3}}\right),$$

and, again, the sum in the exponential is convergent, so $B^\gamma \geq e^{-c\gamma}$ for some absolute constant $c$. □

## 5 Acknowledgements

# References

[1] Richard Arratia, Béla Bollobás, Don Coppersmith, and Gregory B. Sorkin. Euler circuits and DNA sequencing by hybridization. *Discrete Applied Mathematics*, **104**(1-3):63–96, 2000.

[2] Edward A. Bender and E. Rodney Canfield. The Asymptotic Number of Labeled Graphs with Given Degree Sequences. *Journal of Combinatorial Theory, Series A*, **24**(3):296–307, 1978.

[3] Béla Bollobás. A probabilistic proof of an asymptotic formula for the number of labelled regular graphs. *European Journal of Combinatorics*, **1**:311–316, 1980.

[4] Charles J. Colbourn, Wendy J. Myrvold, and Eugene Neufeld. Two Algorithms for Unranking Arborescences. *Journal of Algorithms*, **20**(2):268–281, 1996.

[5] Alan M. Frieze, Mark Jerrum, Michael Molloy, Robert W. Robinson, and Nicholas C. Wormald. Generating and Counting Hamilton Cycles in Random Regular Graphs. *Journal of Algorithms*, **21**(1):176–198, 1996.

[6] Svante Janson. Random Regular Graphs: Asymptotic Distributions and Contiguity. *Combinatorics, Probability & Computing*, 4:369–405, 1995.

[7] Svante Janson, Tomasz Łuczak, and Andrzej Riciński. *Random Graphs*. Wiley, 2000.

[8] Brendan D. McKay and Robert W. Robinson. Asymptotic Enumeration of Eulerian Circuits in the Complete Graph. *Combinatorics, Probability & Computing*, **7**(4):437–449, 1998.

[9] Michael S. O. Molloy, Hanna D. Robalewska, Robert W. Robinson, and Nicholas C. Wormald. 1-Factorizations of random regular graphs. *Random Structures and Algorithms*, **10**(3):305–321, 1997.

[10] James Gary Propp and David Bruce Wilson. How to Get a Perfectly Random Sample From a Generic Markov Chain and Generate a Random Spanning Tree of a Directed Graph. *Journal of Algorithms*, **27**:170–217, 1998.

[11] Hanna D. Robalewska. 2-factors in random regular graphs. *Journal of Graph Theory*, **23**(3):215–224, 1996.

[12] Robert W. Robinson and Nicholas C. Wormald. Almost All Cubic Graphs Are Hamiltonian. *Random Structures & Algorithms*, **3**(2):117–126, 1992.

[13] Robert W. Robinson and Nicholas C. Wormald. Almost All Regular Graphs Are Hamiltonian. *Random Structures & Algorithms*, **5**(2):363–374, 1994.

[14] C. A. B. Smith and W. T. Tutte. On unicursal paths in networks of degree 4. *American Mathematical Monthly*, **48**:233–237, 1941.

[15] Richard P. Stanley. *Enumerative Combinatorics*, volume 2. Cambridge University Press, 2001.

[16] W.T. Tutte. *Graph Theory, Encyclopedia of Mathematics and it Applications*, volume 21. Addison-Wesley Publishing Company, 1984.

[17] Tanja van Aardenne-Ehrenfest and Nicholas G. de Bruijn. Circuits and trees in oriented linear graphs. *Simon Stevin*, **28**:203–217, 1951.

[18] Virginia Vassilevska Williams. Breaking the Coppersmith-Winograd barrier. Manuscript, 2011.